\documentclass[twocolumn,showpacs,preprintnumbers,prl]{revtex4}

\usepackage{graphicx}% Include figure files
\usepackage{dcolumn}% Align table columns on decimal point
\usepackage{bm}% bold math

\begin{document}

\title{Multiple-Retrapping Process in High-$T_c$ Intrinsic Josephson Junctions}

% Force line breaks with \\
\author{Myung-Ho Bae$^{1,\ast}$}
\author{M. Sahu$^1$}
\author{ Hu-Jong Lee$^{2}$}
\author{A. Bezryadin$^{1,\dag}$}
\affiliation{$^1$Department of Physics, University of Illinois at
Urbana-Champaign, Urbana, Illinois 61801-3080, USA}
\affiliation{$^2$Department of Physics, Pohang University of Science
and Technology, Pohang 790-784, Republic of Korea}%
\date{\today}

\begin{abstract}
We report measurements of switching-current distribution (SWCD) from
a phase-diffusion branch to a quasiparticle-tunneling branch as a
function of temperature in a cuprate-based intrinsic Josephson
junction. Contrary to the thermal-activation model, the width of the
SWCD increases with decreasing temperature, down to 1.5 K. Based on
the multiple-retrapping model, we quantitatively demonstrate that
the quality factor of the junction in the phase-diffusion regime
determines the observed temperature dependence of the SWCD.
\end{abstract}


\maketitle

%\section{Introduction}

The escape of a system trapped in a metastable state governs the
reaction rate in various dynamical systems, where the escape is made
by a noise-assisted hopping \cite{Hanggi}. In the case of a
Josephson junction (JJ), a thermal noise induces an escape of a
phase particle representing the system from a local minimum of the
potential well. In an underdamped JJ with hysteresis in the
voltage-current ({\it V-I}) characteristics a single escaping event
induces a switching from a zero-voltage phase-trapped state to a
finite-voltage phase-rolling state \cite{Fulton}. In an overdamped
JJ without hysteresis, however, the energy of an escaped phase
particle is dissipated during its motion so that the particle is
retrapped in another local minimum of the potential. The phase
particle repeats this thermally activated escape and retrapping
process, {\it i.e.}, the multiple-retrapping process. It results in
a phase-diffusion branch (PDB); a resistive branch with a small but
finite voltage for a bias current below a switching current,
$I_{SW}$ \cite{Tinkham}.

A hysteretic JJ can also evolve into this multiple-retrapping state
with increasing temperature ($T$) when the energy fed by a bias
current, near a mean switching current that is suppressed by an
increment in $T$, gets comparable to the dissipated energy.
Recently, this phase-retrapping phenomenon in the hysteretic JJ has
been intensively studied in association with the $T$ dependence of a
switching current distribution (SWCD) \cite{Kivioja,Krasnov}. The
main finding of these studies is that the retrapping process in the
hysteretic JJ modifies the switching dynamics, in such a way as to
\emph{reduce the width of SWCD with increasing $T$}. This SWCD
behavior, contrasting to the usual thermal activation model, has
also been suggested to be caused by the enhanced dissipation due to
a high-frequency impedance, which originates from the contribution
of the impedance of the measurement lines \cite{Martinis,Johnson},
incorporating into temperature effect \cite{Mannik}. At sufficiently
high $T$, the hysteretic JJ begins to show a PDB. This PDB in a
hysteretic JJ can be found whenever the thermal energy, $k_BT$, is
comparable to the Josephson energy, $E_J$ (=$\hbar I_c/2e$; $I_c$ is
the noise-free critical current and $k_B$ is the Boltzmann
constant). Thus, an ultrasmall hysteretic JJ exhibits the PDB even
at a-few-mK region \cite{Vion}. As the dissipation is represented by
the quality factor $Q$ of the JJ, however, it has not been clearly
resolved yet how $Q$, depending both on the temperature and the
impedance, affects switching between a PDB and a
quasiparticle-tunneling branch (QTB).

In this letter, we study switching from the PDB to the QTB in
intrinsic Josephson junctions (IJJs) of
Bi$_2$Sr$_2$CaCu$_2$O$_{8+x}$ (Bi-2212) at various temperatures. The
effective junction impedance $Z_J$ including the influence of an
external circuit in a PDB, was estimated by fitting {\it V-I} curves
to the phase-diffusion model. The values of $Z_J$ turn out to be an
order of the measurement-line impedance $Z_L$, which indicates that
the phase dynamics of our IJJs is governed by the environmental
dissipation. Our measurements of the $T$ dependence of the switching
rate $\Gamma_S$ and the corresponding SWCD in a single IJJ are in
good agreement with those estimated by the multiple-retrapping
model. This study quantitatively clarifies how the impedance- and
temperature-related dissipations, and the corresponding $Q$
determine the switching dynamics in a hysteretic IJJ with
phase-diffusion characteristics.

A stack with the lateral size of 2.5$\times$2.9 $\mu$m$^2$ in
Bi-2212 single crystal was defined by using focused-ion-beam (FIB)
process \cite{Kim}. High-intensity FIB irradiation is known to
degrade the peripheral region by the scattered secondary ion beam
\cite{Bae}. In the milling process, we used a relatively high
ion-beam current of 3 nA, corresponding to $\sim$200 pA/$\mu$m$^2$.
This high ion-beam current much reduced the interlayer tunneling
critical current density down to $\sim$8 A/cm$^2$ in $N$=12
junctions out of total $\sim$100 junctions in a stack, which were
estimated from the number of QTBs in {\it V-I} curves of the stack.
Four-terminal transport measurements were carried out in a pumped
He$^4$ dewar with the base temperature of 1.45 K. Room-temperature
$\pi$-filters were employed and measurement lines were embedded in
silver paste at cryogenic temperatures to suppress high-frequency
noises propagating along the leads. The measurements were made by
using battery-operated low-noise amplifiers (PAR-113). The ramping
speed of the bias current and the threshold voltage (obtained from
the maximum voltage of the PDB) in measuring $I_{SW}$ were
$\dot{I}$=30 $\mu$A/sec and $V_{th}$=110 $\mu$V \cite{Vth},
respectively. Measurements were made on the first jump, which
connected the PDB to the first QTB of the weakest junction in the
stack. For each distribution, 10000 switching events were recorded
with the current resolution of 90 pA.

\begin{figure}[t]
\begin{center}
\leavevmode
%h=here, t=top, b=bottom, p=separate figure page
\includegraphics[width=.8\linewidth]{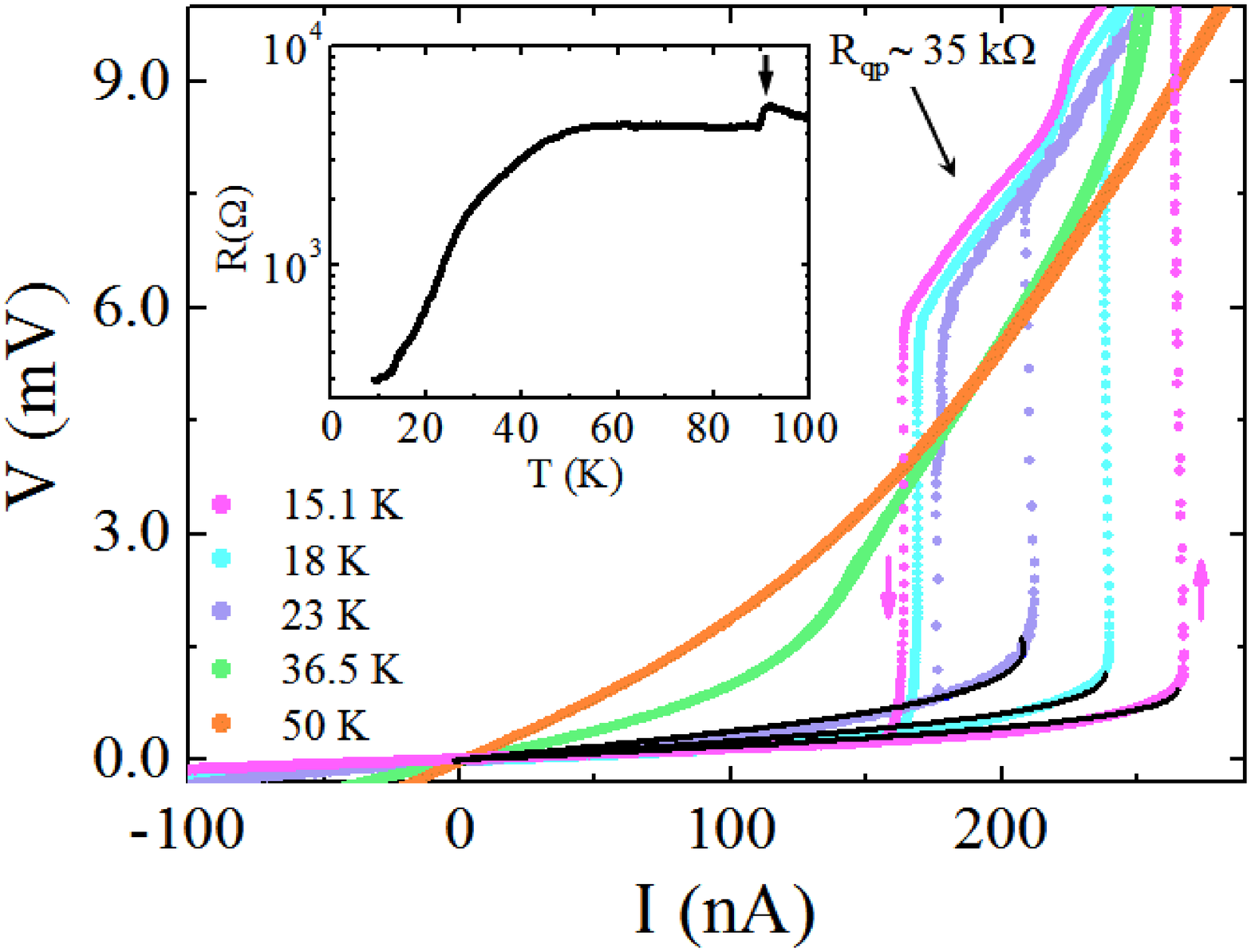}
\caption{(color online) {\it V-I} characteristics at various
temperatures of the Bi-2212 IJJs stack. The upward and downward
arrows indicate the switching and the return currents, respectively,
at $T$=15.1 K. Black solid curves are the best fits to the phase
diffusion model at three different temperatures of 15.1 K, 18 K, and
23 K, with the effective junction impedance $Z_J$=125, 125, and 128
$\Omega$ as the best-fit parameters, respectively. Inset: the
$R$-vs-$T$ curve, where $T_c$ is indicated by an arrow.}
\end{center}
\end{figure}

The inset of Fig. 1 shows the tunneling $R$ vs $T$ curves with the
superconducting transition at $T_c$=90 K, which is indicated by an
arrow. The $c$-axis tunneling nature in the Bi-2212 stacked
junctions is evident by the increasing resistance with decreasing
$T$ above $T_c$. At $T$$<$$T_c$, the junction resistance does not
vanish completely, even down to $T$=10 K, which is attributed to the
phase diffusion in the junctions. Fig. 1 shows the {\it V-I} curves
at various $T$ below $T_c$. The nonlinear curve at $T$=50 K begins
to show a hysteresis below $T$$\sim$30 K. For a bias below a
$I_{SW}$ in each curve, a pronounced non-zero resistive branch
appears below $T$=23 K. Here, the finite resistance is caused by the
phase diffusion in the condition of $k_BT$$\sim$$2E_J$
\cite{Tinkham}. The phase-diffusion model \cite{Ingold,Franz}
predicts the current-voltage relation of
$I(V)=\frac{4eI_{SW}k_BT}{\hbar}\frac{Z_Jv}{v^2+[2eZ_Jk_BT/\hbar]^2}$
with a frequency-dependent effective junction impedance $Z_J$. Here,
the phase diffusion is assumed to take place in all twelve junctions
with FIB-suppressed critical currents, so that the voltage bias per
junction becomes $v=V/N$. We use this relation to fit the PDB as
shown by black curves in Fig. 1 with $Z_J$ as the best-fit
parameter. $Z_J$ turns out to be 125, 125, and 128 $\Omega$ for
$T$=15.1, 18, and 23 K, respectively. These values of $Z_J$ are
comparable to the measurement line impedance $Z_L$ (50$\sim$100
$\Omega$), even with a much higher quasiparticle tunneling
resistance of $R_{qp}$$\sim$35 k$\Omega$ as seen in Figure 1. This
implies that the dissipation in the PDB is dominated by the
high-frequency (of an order of the Josephson plasma frequency,
$\omega_p$) dissipation through the measurement lines \cite{Johnson,
Martinis}. Since the energy of the escaped phase particle is
dissipated through the environment in the phase-diffusive regime,
the switching also becomes sensitive to this dissipation process.

Now, we turn to the switching event from the PDB to the QTB. Fig.
2(a) shows the SWCD (scattered symbols) at various temperatures. The
standard deviation ($\sigma$) and the mean switching currents
($\langle I_{SW}\rangle$) are shown as a function of $T$ in the
inset of Fig. 2(a). The $T$ dependence of $\sigma$ contradicts to
that of a conventional underdamped JJ, where $\sigma$ increases with
$T$ in a thermal-activation regime \cite{Fulton}. Fig. 2(b) shows
$\Gamma_S$ (scattered points) vs $I$ calculated from the SWCD of
Fig. 2(a) following the Fulton and Dunkleberger analysis
\cite{Fulton}. With lowering $T$, $\Gamma_S$($I$) shows a pronounced
change from an almost linear to a down-turn nonlinear bias-current
dependence in a semi-logarithmic plot. To explain this behavior, we
adopted the multiple-retrapping model \cite{Mannik,Krasnov}. In the
phase-diffusive regime, the successive retrapping processes suppress
the switching rate, $\Gamma_S$. The switching to the high-voltage
branch (i.e. the QTB) occurs only when the phase particle is not
retrapped after escaping from a local energy minimum. A phase
particle escaped from a potential minimum has a probability,
$P_{RT}$, to be retrapped in the next potential minimum. The
switching rate $\Gamma_S$, including $P_{RT}$, is expressed by
\cite{Mannik}
\begin{equation}
\Gamma_S=\Gamma_{TA}(1-P_{RT})\frac{\mbox{ln}(1-P_{RT})^{-1}}{P_{RT}}.
\end{equation}
Here,
$\Gamma_{TA}$[=$\frac{\omega_p}{2\pi}\mbox{exp}($-$\frac{\Delta
U}{k_BT})]$ is the thermally activated escape rate,
$\omega_p$=$\omega_{p0}(1$-$\gamma^2)^{1/4}$,
$\omega_{p0}$=$(2eI_c/\hbar C)^{1/2}$, $\Delta
U(\phi)$[=$2E_J((1$-$\gamma^2)^{1/2}$-$\gamma\mbox{arccos}\gamma)]$
is the escape energy barrier, and $\gamma$[=$I/I_c]$ is a normalized
bias current. The retrapping probability can be obtained by an
integration (see Ref. \cite{Garg}) of the retrapping rate
$\Gamma_{RT}$=$\frac{1}{Z_JC}(\Delta U_{RT}/\pi
k_BT)^{1/2}\mbox{exp}(-\Delta U_{RT}/k_BT)$, where $\Delta
U_{RT}(I)$=$Z_J^2C(I$-$I_{r0})^2$ and $I_{r0}$ is the noise-free
return current from a QTB to a PDB \cite{Ben}. Fig. 3(a) shows the
experimental switching distribution (red dots) at $T$=1.5 K with the
corresponding fit (green curve) obtained by using Eq. (1) with the
best-fit parameters of $Z_J$=61.9 $\Omega$, $I_c$=1.26 $\mu$A, and
$I_{r0}$=63 nA. The junction capacitance, 330 fF, was estimated from
the typical value of 45 fF/$\mu$m$^2$ for Bi-2212 IJJs \cite{Irie}.
The corresponding switching rate (red dots) and the fit (green
curve) are shown in Fig. 3(b), with the same parameters. An
excellent agreement is obtained in both fittings.

\begin{figure}[t]
\begin{center}
\leavevmode
%h=here, t=top, b=bottom, p=separate figure page
\includegraphics[width=0.84\linewidth]{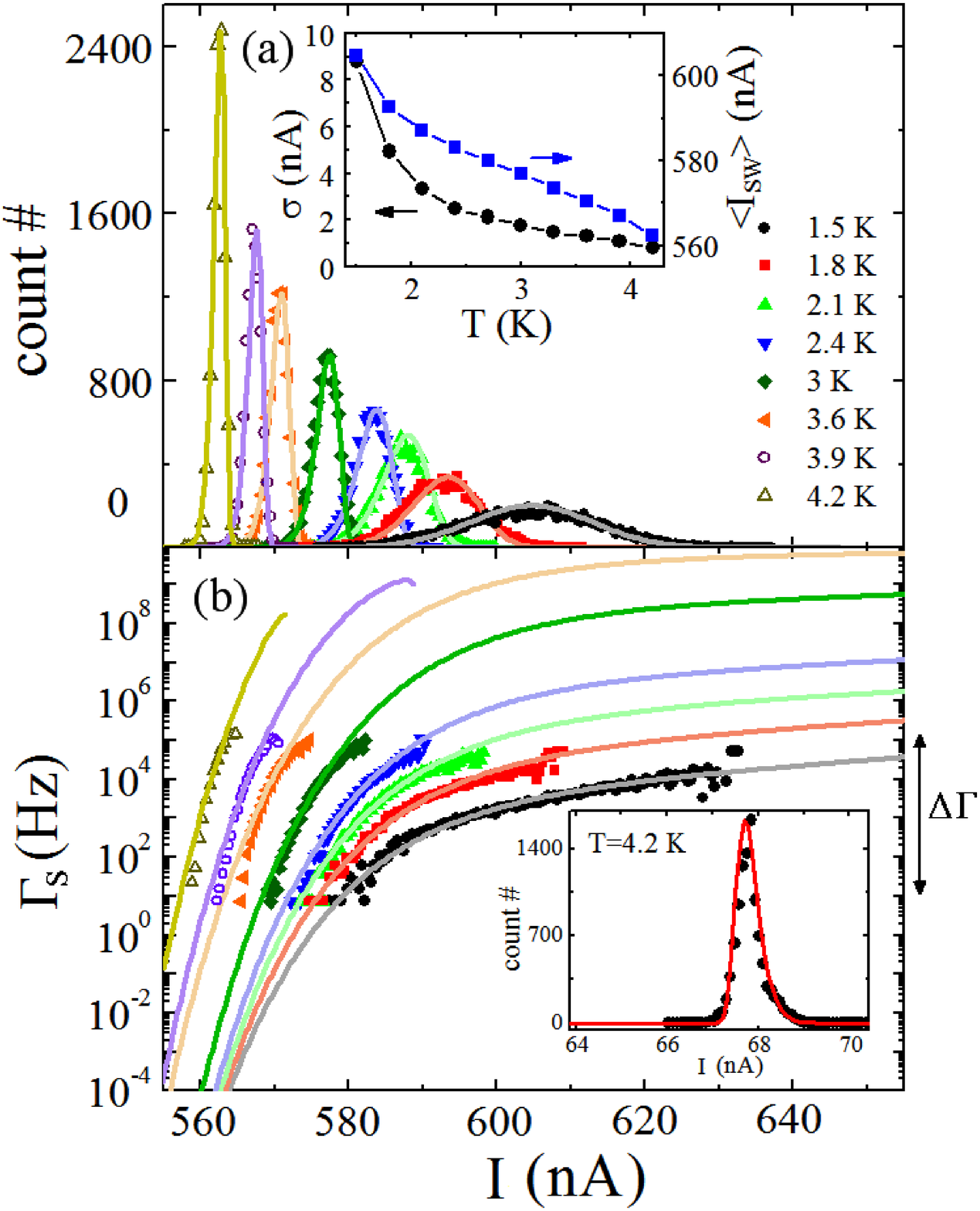}
\caption{(color online) (a) The switching current distribution and
(b) the corresponding  $\Gamma_S$ at various $T$ from 1.5 K to 4.2
K. The solid lines are calculated based on the multiple-retrapping
model. Inset: Temperature dependence of the standard deviation
($\sigma$) and the mean switching current ($\langle I_{SW}\rangle$).
$\Delta\Gamma$ is the range of the observable switching rate for a
given $\dot{I}_b$. Inset: the retrapping current distribution (solid
circles) and the calculated behavior (solid line) at $T$=4.2 K.}
\end{center}
\end{figure}

\begin{figure}[t]
\begin{center}
\leavevmode
%h=here, t=top, b=bottom, p=separate figure page
\includegraphics[width=0.84\linewidth]{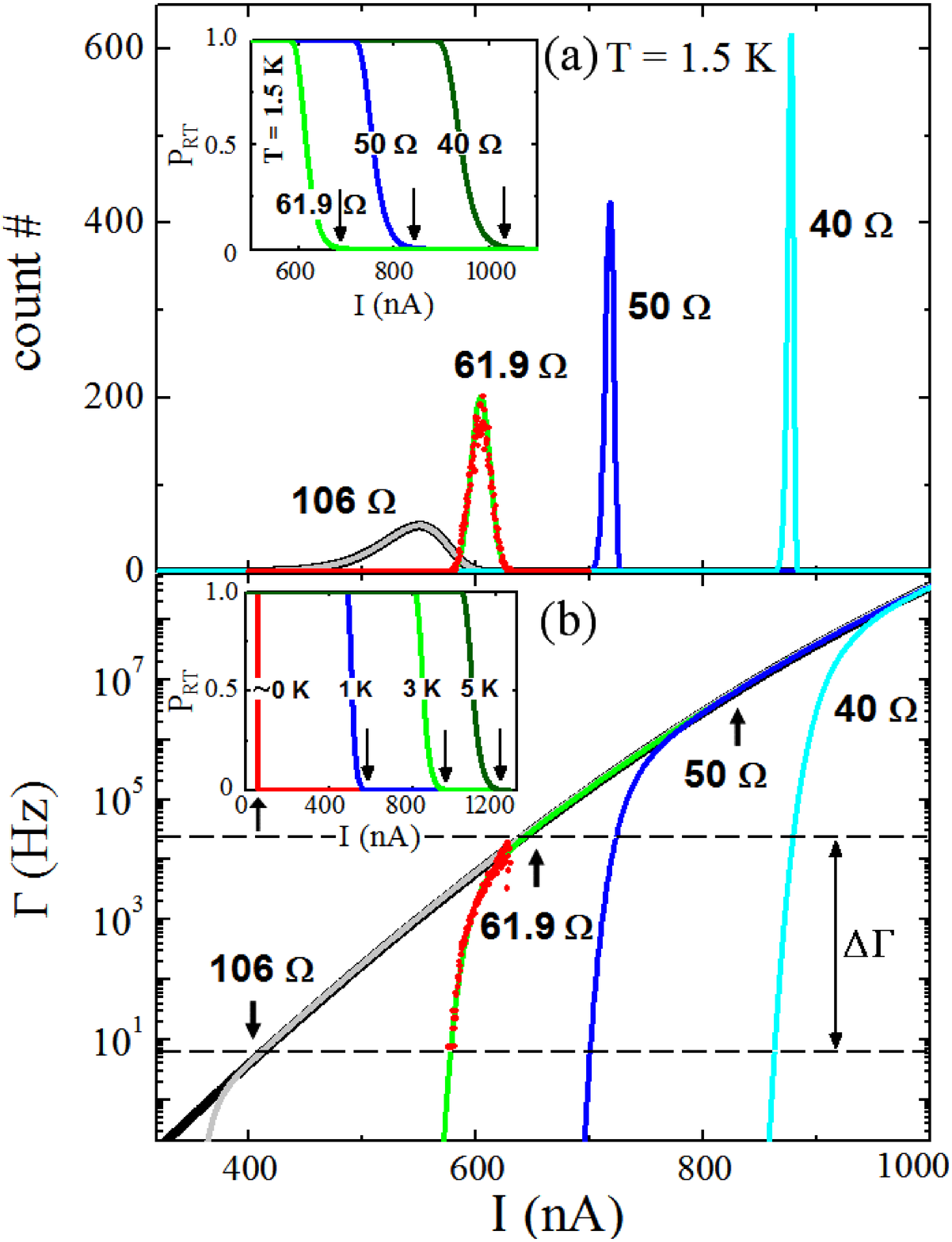}
\caption{(color online) (a) The solid lines represent the calculated
switching current distributions for various $Z_J$ values of 106,
61.9, 50, and 40 $\Omega$, corresponding to $Q_{PD}$=3.88, 2.37,
1.90 and 1.55, respectively. Red dots show the SWCD at $T$=1.5 K.
Inset: Theoretical retrapping probabilities versus bias current, for
various $Z_J$ values at $T$=1.5 K. The downward arrows indicate the
$I_{PD}$ values. (b) Estimated switching rate versus the bias
current. Red dots correspond to the experimental switching rate for
$T$=1.5 K. Estimated $\Gamma_S$ for various $Z_J$ values in Eq. (1)
are shown as solid lines. Inset: Estimated retrapping probabilities
versus bias current with $Z_J$=61.9 $\Omega$ at various $T$. Thick
black curves in (a) and (b) show the thermally activated switching
current distribution and the corresponding escape rate,
respectively, without retrapping.}
\end{center}
\end{figure}

%To these results we need to consider $Z_J$- and
%$T$-dependence of $P_{RT}$.
These results are analyzed in terms of $Z_J$- and $T$-dependence of
$P_{RT}$. The inset of Fig. 3(a) shows the calculated
$P_{RT}$-vs-$I$ curves for various $Z_J$ at $T$=1.5 K, with the
values of $I_c$ and $I_{r0}$ obtained from the best fits of Figure
3. The retrapping-probability curve shifts to higher currents as
$Z_J$ decreases, due to the presence of a $Z_J$ dependence of
$\Delta U_{RT}$ in the exponential factor of $\Gamma_{RT}$. The
current positions of almost vanishing $P_{RT}$, indicated by
downward arrows, approximately correspond to the maximum current
allowing the retrapping. We denote this current as $I_{PD}$.
Physically this is the same current as the one denoted $I_m$ in Ref.
\cite{Martinis}. The system can hardly be retrapped at a current
higher than $I_{PD}$, because in this case the energy fed to the
system by the bias current gets larger than the dissipated energy.
By equating the energy fed and the energy dissipated, similar to
McCumber and Stewart analysis \cite{Kivioja}, one obtains the
relation
\begin{equation}
I_{PD}=4I_c/\pi Q_{PD},
\end{equation}
where $Q_{PD}$ is the phase-diffusion quality factor at
$\omega$$\sim$$\omega_p$ \cite{Martinis}. In fact, the noise-free
retrapping current can be written %similarly
in a form similar to Eq. (2), namely as $I_{r0}=4I_c/\pi
Q(\omega$=0) \cite{Tinkham}. The thick black curves in Figs. 3(a)
and 3(b) show SWCD and the corresponding $\Gamma_{TA}$($I$),
respectively. Other solid curves in Fig. 3(b) are $\Gamma_S$($I$) in
Eq. (1) for varying $Z_J$ under the multiple retrapping processes.
$\Gamma_S$($I$) with each $Z_J$ in Fig. 3(b) starts to drop quickly
from $\Gamma_{TA}$($I$) at $I$=$I_{PD}$, which are denoted by arrows
in the figure. We define $I_{PD}$ as the current value corresponding
to $P_{RT}$=0.01, where $\Gamma_S(I_{PD})$ is nearly the same as the
$\Gamma_{TA}(I_{PD})$ as shown in Fig. 3(b) \cite{Mannik}. The
impedance of 106 $\Omega$ gives the same SWCD as thermally activated
SWCD without retrapping because $\Gamma_S(I)$ in $\Delta\Gamma$ well
overlaps with $\Gamma_{TA}(I)$ although the impedance is of an order
of $Z_L$. When $\Gamma_S$ $(I_{PD})$ crosses over the bottom line of
$\Delta\Gamma$ while $Z_J$ keeps decreasing, the high-frequency
dissipation affects the SCDW: the observable window of $\Delta
\Gamma$ (between the two dashed lines) for a fixed $\dot{I}_b$
shifts to the steeper section with decreasing $Z_J$, resulting in
the decreasing width of the SWCD in Figure 3(a) at a constant $T$.
This shift itself is caused by a reduction of $Q_{PD}$ with
decreasing $Z_J$ at a constant $T$.

Since $I_{PD}$ is also affected by a change in $T$ as followings,
the shape of $\Gamma_S$($I$) dose not simply follow the variation of
$Z_J$ as $T$ changes. The inset of Fig. 3(b) illustrates $P_{RT}$ vs
$I$ at various temperatures. Here, $Z_J$ is fixed at 61.9 $\Omega$
and other parameters, except for $T$, are set to be the same as for
the inset of Fig. 3(a). The current position of the zero-temperature
curve, indicated by an upward arrow in the inset of Fig. 3(b),
corresponds to the fluctuation-free return current $I_{r0}$. The
value of $I_{PD}$ shown by arrows increases with increasing $T$.
Thus, in effect, the retrapping probability curve shifts to higher
currents as $T$ is raised, due to the presence of a Boltzmann-type
exponential factor in the expression of $\Gamma_{RT}$. Fig. 2
illustrates the calculated SWCD and $\Gamma_S$($I$) as solid curves
at various temperatures with the best-fit parameters listed in Table
I, which agree well with the observation. Here, since the
dissipation effect by temperature is already determined by a bath
temperature, the main fit-parameter becomes $Z_J$ in Figure 3(a). As
shown in Fig. 2(b), with increasing $T$, the calculated
$\Gamma_S$$(I)$ in the window of $\Delta\Gamma$ shows steeper
regions. It causes the reduction of the width of SWCD with
decreasing temperature. The ratio of $I_c$ and $I_{PD}$ becomes
smaller with increasing $T$ as in Table I. $I_{PD}$ even equals to
$I_c$ at $T$=4.2 K. This behavior leads to the conclusion that
$Q_{PD}$ decreases with increasing $T$ following Eq. (2) despite the
increase of $Z_J$ with $T$ \cite{ZJ}. The slop of the calculated
$\Gamma_S$($I$) in the window of $\Delta\Gamma$ at sufficiently high
temperatures is insensitive to $T$ variations, which leads to the
saturation of $\sigma$ with increasing $T$ as the inset of Figure
2(a).

\begin{table}
\caption{\label{tab:tableI}Fitting parameters for the switching
events for selected temperatures}
\begin{ruledtabular}
\begin{tabular}{ccccccc}
$T$(K)&$I_c$($\mu$A)&$I_{r0}$(nA)&$Z_J$($\Omega$)&$I_{PD} (nA)$&
$Q_{PD}$&\\
\hline
1.5 &1.263&63.00&61.9&678&2.37\\
2.4 &1.209&65.10&77.6&683&2.25\\
3.0 &1.006&63.44&86.0&685&1.87\\
3.6 &0.735&65.22&94.7&682&1.37\\
4.2 &0.572&63.79&101.5&572&1.27\\
\end{tabular}
\end{ruledtabular}
\end{table}

Now we consider the retrapping dynamics related to the transition
from a QTB to a PDB in an underdamped JJ. It can be shown that
only the zero-frequency dissipation plays an significant role in
the retrapping dynamics \cite{Martinis}. The inset of Fig. 2(b)
shows the stochastic nature of the retrapping current as shown by
solid circles at $T$=4.2 K. The calculated retrapping distribution
(solid line) by the retrapping rate, $\Gamma_{RT}$, is consistent
with the experimental result. The best-fit parameters are
$I_{r0}$=63.8 nA and $Z_J$=10 k$\Omega$ ($\sim$$R_{qp}$), where
the estimated noise-free $I_{r0}$ matches with the value used in
$\Gamma_S$ fitting in Table I. The junction impedance $Z_J$
estimated from the return currents is significantly larger than
$Z_J$ ($\omega$$\sim$$\omega_p$) for the observed switching
events. It indicates that the retrapping phenomena from QTB to PDB
are mainly determined by a zero-frequency damping with $Q$(0)=11.4
with $I_c$=572 nA and $I_{r0}$=63.8 nA at $T$=4.2 K as shown in
Table I.

In summary, we clearly show that the multiple retrapping processes
in underdamped IJJs, with much suppressed critical current and high
tunneling resistance, govern the switching from the PDB to the QTB.
The predicted SWCD and $\Gamma_S$ in the multiple-retrapping model
are in good agreement with the observed broadening of the
distribution of switching currents with decreasing temperature. We
also demonstrate that the change of the shapes of the observed SWCD
and the $\Gamma_S$ in various temperatures can be understood by
impedance and temperature dependence of $Q_{PD}$, taking the
retrapping probability into account. As the macroscopic quantum
tunneling has been observed recently in IJJs
of Bi-2212 single crystals \cite{MQT}, this study %could
provides in a quantitative manner the role of the dissipation in
quantum devices based on cuprate-based JJs.

This work was supported by DOE Grants No. DEFG02-07ER46453. We
acknowledge the access to the fabrication facilities at the
Frederick Seitz Materials Research Laboratory. This work was also
partially supported by POSTECH Core Research Program, Acceleration
Research Grant No. R17-2008-007-01001-0, and the Korea Research
Foundation Grants No. KRF-2006-352-C00020.

$^\ast$mhbae@uiuc.edu, $^\dag$bezryadi@uiuc.edu


\begin{thebibliography}{00}
\bibitem{Vth} In our measuremet setup, the voltage
spacing beween the maximum voltage value of the PDB and the next
jumping voltage is $\sim$150 $\mu$V so that the
value of our $V_{th}$ of 110$\mu$V is reasonable %value
to obtain $I_{SW}$ values.
\bibitem{ZJ} The noise-free critical current, $I_{c}$, is strongly suppressed with
$T$, resulting in a reduction of $\omega_p$. Thus, the frequency
related to a phase diffusion dynamics gets smaller, leading to a
larger value of $Z_J$ with $T$ as shown in Table I \cite{Martinis}.
\end{thebibliography}
\end{document}